\documentclass{article}

\usepackage{amssymb}
\usepackage{txfonts}
\usepackage{graphicx}
\usepackage{bm}
\usepackage{subeqn}

\begin{document}

\title{Fractional Calculus Approach to the Statistical Characterization of Random Variables and Vectors}
\author{Giulio Cottone$^1$, Mario Di Paola$^1$ and Ralf Metzler$^2$ \thanks{e-mails: giuliocottone@unipa.it, mario.dipaola@unipa.it,metz@ph.tum.de}\\
\small$^1$ Dipartimento di Ingegneria Strutturale, Aerospaziale e 
 Geotecnica\\ \small Universit\'a degli Studi di Palermo, Viale delle 
 Scienze, 90128 Palermo, Italy\\
\small
$^2$ Physik Department, Technical University of Munich, 85748 Garching, Germany
}

\date{}

\maketitle

\begin{abstract}
Fractional moments have been investigated by many authors to represent the density of univariate and bivariate random variables in different contexts. Fractional moments are indeed important when the density of the random variable has inverse power-law tails and, consequently, it lacks
integer order moments. In this paper, starting from the Mellin transform of the characteristic
function and by fractional calculus method we present a new perspective on the statistics of random variables. 
Introducing the class of complex
moments, that include both integer and fractional moments, we show
that every random variable can be represented within this approach,
even if its integer moments diverge.
Applications to the statistical characterization of raw data and in the representation of both random variables and vectors are provided, showing that the good numerical convergence makes the  proposed approach a good and reliable tool also for practical data analysis. 
\end{abstract}

\section{Introduction}
Fractional moments are very useful in dealing with random variables
with power-law distributions, $F(x)\sim\left| x \right|^{ - \mu } $, $\mu  >
0$, where $F(x)$ is the distribution function. Indeed, in such cases, moments $\left\langle {\left| X \right|^q } \right\rangle $ exist only if $ q < \mu $ and integer order moments greater than $\mu$ 
diverge. Distributions of this type are encountered in a wide
variety of contexts, see the extensive literature in \cite{metz00} and \cite{metz04}
where power-law statistics appear in the framework of anomalous diffusion
in many fields of applied science.
To name but a few examples, we mention the travel length distribution
in human motion patterns and animal search processes
\cite{brockmann,barabasi,lomholt,mateos}, fluctuations in plasma devices
\cite{chechkin,diego}, scaling laws in polymer physics and its applications
to gene regulation models \cite{duplantierreview,lomholt1}, as well as the
distribution of time scales in processes such as the motion of charge
carriers in amorphous semiconductors \cite{scher}, tracer dispersion in
groundwater \cite{scher1}, or sticking times in turbulent flows \cite{weeks}.

The lack of moments is of course a great limitation for the characterization
of such distributions, because many methods of statistical analysis fail,
like for example, the representation of the characteristic function or the
log-characteristic function by moments or by cumulants, respectively. Then,
it is necessary to resort to other distribution properties and a good option
is given by so-called fractional moments which have been extensively
studied in the seventies and recently attracted new interest both
in theoretical \cite{gzyl06,novi03,novi05} and in
experimental settings \cite{nigm09}. In particular, in \cite{novi03}
and \cite{novi05} the moment problem originally stated in terms of integer
moments, has been extended to fractional moments where it has been shown
that the knowledge of some fractional moments $\langle {X^q}\rangle$
improves significantly the convergence speed of the maximum entropy
method for \textsl{non-negative} densities i.e. defined in $[0,\infty[$,
if a proper optimization procedure is used in selecting the order $q$. 
This can be explained by the non-local nature of fractional moments that will be better highlighted in the following of the paper.

Fractional moments of a non-negative random variable are expressible by
Mellin transform of density and this fact has been widely used in literature
principally in the field of the algebra of random variables. That is, Mellin
transform is the principal mathematical tool to handle problems involving
products and quotients of independent random variables. As the approach
presented in this paper is based on Mellin transform, a brief outline of
previous results is given in the following.

The first author who developed a systematic method to express the density of
the product of independent random variables by Mellin transform was Epstein. As
consequence, in his pioneering work \cite{epst48} he recognizes that any
density function can be obtained as Mellin convolution of two independent
densities. It must be remarked that such a result was already obtained
by \cite{hunt39} outside the framework of integral transforms. Many other
studies on the use of Mellin transform extending Epstein's work have been
reported providing explicit analytic forms for products of independent random variables
with assigned densities. In \cite{dola64} examples on product and quotient
of independent random variables with Rayleigh distribution and moments of
Rice distribution are obtained by Mellin transform tables. An extensive
use of such concepts and the state-of-art on the algebra of random variable
may be found in the the book of \cite{spri79}.

Another research direction on Mellin applications in probability is represented
by the use of special functions like the H-Fox and the Maijer's G-functions,
due principally to Mathai and co-workers. Such functions are indeed
representable as Mellin-Barnes integrals of the product of gamma functions
and are therefore suited to represent statistics of products and quotients of
independent random variables whose fractional moments are expressible as gamma
or gamma related functions. Applications of special functions to statistics
and probability theory can be found in \cite{math93b, math08, math78,
math93, prov86, prov88} and references therein.

Whether fractional moments exists and how they are correlated to local
properties of the characteristic function has been investigated
in \cite{wolf75, wolf71, wolf73, wolf75b}. In
particular, in \cite{wolf75} a relation is derived between the absolute
moment of real order and integrals involving the Marchaud fractional derivative
of the characteristic function. Other relevant studies on the existence
of fractional (absolute) moments can also be found in \cite{laue80,
laue83, laue86}.

All the cited work share the common issue of using the Mellin transform because it naturally coincides with moments of the type $ \langle X^{\gamma-1} \rangle$ if the random variable $X$ has a density defined only in the positive domain. 
In this paper, following the previous results in \cite{cott09}, it is shown that applying the Mellin transform to the \textit{characteristic function}, and not to the density, a sound representation of the statistics of general multivariate random variables is possible. In this way, exploiting the Hermitian nature of the characteristic function, a unique representation valid for every random variable and vectors, defined in the whole real range is presented. 
Further, by means of simple properties of fractional operators, a generalization of the well-known link between moments and the derivative of the characteristic function evaluated in zero is provided. Such a link is useful because it allows one to find existence criteria for fractional moments that can be derived in a simpler way with respect to \cite{wolf75} and clarifies the non-local nature of fractional moments. Moreover, this approach that starts from the characteristic function plainly applies to random variables defined in the whole domain without the need of partitioning. Such theoretical aspects  go along with a better numerical treatment of the inverse Mellin transform involved in this approach as it will stressed in the following. 

In the application section,
it will be indeed shown that the proposed approach is extremely useful in the statistical characterization of raw data and in the representation of both random variables and vectors with few computational effort. Thus, the good numerical convergence makes the approach proposed a good and reliable tool also for data analysis. 
We want to stress that the representation proposed in this paper can be applied to any Fourier pair functions used in statistical analysis and, very recently, applications to the representation of stochastic processes have been proposed in \cite{cott09b, cott09c}.

\section{Probabilistic Characterization of Random Variables}
A brief outline of the basic tools of probability will serve to frame the problem in hand and to introduce symbols. Let $\left( {{\cal S},{\cal A},{\cal P}} \right)$ be a probability space, where ${\cal A}$ is a collection of events defining a $\sigma $-algebra of subsets on the sample space ${\cal S}$, and ${\cal P}$ a probability measure that assigns a number to each event on ${\cal A}$ and $X:\left( {{\cal S},{\cal A}} \right) \to \left( {\mathbb{R},{\cal B}} \right)$ a random variable, where ${\cal B}$ is a Borel set. $F\left( x \right) = \Pr \left( {X < x} \right)$ is the cumulative distribution and $p\left( x \right) = {\rm{d}} F/{\rm{d}} x$ is the probability density function (PDF). The Fourier transform of $p(x)$, denoted as ${\cal{F}} p = \varphi( u )$, is the characteristic function (CF), that is
\begin{equation}
\varphi \left( u \right) = \langle {\exp \left( {{\rm{i}}\,uX} \right)} \rangle = \int_{ - \infty }^\infty  {\exp \left( {{\rm{i}}ux} \right)p\left( x \right) {\rm{d}} x} 
\label{equ1}\\
\end{equation}
with $u \in \mathbb{R}$, ${\rm{i}} = \sqrt { - 1} $ and $\langle  \cdot  \rangle$ means expectation, i.e. $\langle {g\left( X \right)} \rangle = \int_{ - \infty }^\infty  {g\left( x \right)p\left( x \right){\rm{d}} x} $. Expectations of the functions $g\left( X \right) = X^j $ with $j = 1,2,...$, provided they exist, give the integer moments of  $X$, indicated by $\langle {X^j } \rangle$. These integer moments are related to the CF  by the Taylor expansion 
\begin{equation}
\varphi \left( u \right) = \sum\limits_{j = 0}^\infty  \langle {({\rm{i}} X)^j } \rangle{\frac{u^j }{{j!}}}
\label{equ2}\\
\end{equation}
due to the property 
\begin{equation}
\langle {({\rm{i}} X)^j } \rangle = \left. {{\rm{d}}^j \varphi \left( u \right)/{\rm{d}} u^j } \right|_{u = 0} 
\label{equ3}
\end{equation}
 
Whilst eq.(\ref{equ2}) might be useful to evaluate the CF from moments, its Fourier transform that expresses the relation between density and moments is not useful from the computational point of view, as it reads 
\begin{equation}
p\left( x \right) = \sum\limits_{j = 0}^\infty  {\frac{{\left( {{\rm{ - 1}}} \right)^j }}{{j!}}} \langle {X^j } \rangle\frac{{{{\rm{d}}}^j \delta \left( x \right)}}{{{{\rm{d}}}x^j }}
\label{equ4}
\end{equation}
where $\delta \left( x \right)$ is the Dirac's delta. Then, it is not possible to use integer moments to represent the PDF by eq.(\ref{equ4}). It must be stressed that eq.(\ref{equ2}) does not hold for every distribution. This is the case for distributions with heavy tails, and consequently with divergent integer moments. Further, the use of eq.(\ref{equ2}) is not always feasible, although appealing, because truncation of the Taylor expansion involved produces a divergent behavior of the CF at $u \to \infty $. 

In a recent paper \cite{cott09}, it has been proposed to use a new class of moments, namely complex moments, to represent both the CF and the PDF of a random variable.  The fundamentals of the generalization of eq.(\ref{equ2}), eqs.(\ref{equ3}) and (\ref{equ4}) are briefly summarized hereinafter and it will be shown that such representation does not entail the drawback of eqs.(\ref{equ2}) and eq.(\ref{equ4}). 

To this aim, let $\gamma  = \rho  + {\rm{i}}\eta $, with $\rho  > 0$ and $\eta  \in \mathbb{R}$, and let $\left( {I_ \pm ^\gamma  f} \right)\left( x \right)$ and $\left( {{D}_ \pm ^\gamma  f} \right)\left( x \right)$ denote the Riemann-Liouville (RL) fractional integral and derivative, while $\left( {{\bm D}_ \pm ^\gamma  f} \right)\left( x \right)$ is the Marchaud fractional derivative, namely 
\begin{subequations}\label{equ5}
\begin{eqnarray}\label{equ5a}
\left( {I_ \pm ^\gamma  f} \right)\left( x \right)\mathop  = \limits^{def} \frac{1}{{\Gamma \left( \gamma  \right)}}\int_0^\infty  {\xi ^{\gamma  - 1} f\left( {x \mp \xi } \right){\rm{d}} \xi } 
\end{eqnarray}
\begin{eqnarray}\label{equ5b}
\left( {D_ \pm ^\gamma  f} \right)\left( x \right)\mathop = \limits^{def} \frac{1}{{\Gamma \left( {1 - \gamma } \right)}}\frac{{\rm{d}}}{{{\rm{d}}x}}\int_0^\infty  {\xi^{ - \gamma } f\left( {x \mp \xi } \right){\rm{d}}\xi } 
\end{eqnarray}
\begin{eqnarray}\label{equ5c}
\left( {{\bm D}_ \pm ^\gamma  f} \right)\left( x \right)\mathop  = \limits^{def} \frac{1}{{\Gamma \left( { - \gamma } \right)}}\int_0^\infty  {\xi ^{ - \gamma  - 1} \left( {f\left( {x \mp \xi } \right) - f\left( x \right)} \right){\rm{d}}\xi } 
\end{eqnarray}
\end{subequations}
being $\Gamma \left( \gamma  \right)$ the gamma function, and $n$ the integer part of $\rho$+1.

Fourier transform of fractional derivatives, in case $0<\rho<1$, (\cite{samk93}, p.137) gives
\begin{subequations}\label{equ6}
\begin{eqnarray}\label{equ6a}
{\cal F}\left( {I_ \pm ^\gamma  f} \right) = \left( { \mp {\rm{i}}u} \right)^{ - \gamma } {\cal F}f
\end{eqnarray}
\begin{eqnarray}\label{equ6b}
{\cal F}\left( {D_ \pm ^\gamma  f} \right)={\cal F}\left( {{\bm D}_ \pm ^\gamma  f} \right) = \left( { \mp {\rm{i}}u} \right)^\gamma  {\cal F}f
\end{eqnarray}
\end{subequations}
while inverse Fourier transform gives
\begin{subequations}\label{equ7}
\begin{eqnarray}\label{equ7a}
{\cal F}^{ - 1} \left( {I_ \pm ^\gamma  f} \right) = \left( { \pm {\rm{i}}x} \right)^{ - \gamma } {\cal F}^{ - 1} f
\end{eqnarray}
\begin{eqnarray}\label{equ7b}
{\cal F}^{ - 1} \left( {D_ \pm ^\gamma  f} \right)={\cal F}^{ - 1} \left( {{\bm D}_ \pm ^\gamma  f} \right) = \left( { \pm {\rm{i}}x} \right)^\gamma  {\cal F}^{ - 1} f
\end{eqnarray}
\end{subequations}

For functions that are Fourier pairs, the previous equations are very useful for calculating the fractional operators in an easier way with respect to definitions (\ref{equ5}). Indeed, by Fourier transform of eq.(\ref{equ7})
\begin{subequations}\label{equ8}
\begin{eqnarray}\label{equ8a}
\left( {I_ \pm ^\gamma  {\cal F}f} \right)\left( u \right) = \int_{ - \infty }^\infty  {\left( { \pm {\rm{i}}x} \right)^{ - \gamma } \,\exp \left( {{\rm{i}}u x} \right)f\left( x \right){\rm{d}}x}
\end{eqnarray}
\begin{eqnarray}\label{equ8b}
\left( {D_ \pm ^\gamma  {\cal F}f} \right)\left( u \right)=\left( {{\bm D}_ \pm ^\gamma  {\cal F}f} \right)\left( u \right) = \int_{ - \infty }^\infty  {\left( { \pm {\rm{i}}x} \right)^\gamma  \,\exp \left( {{\rm{i}}ux} \right)f\left( x \right){\rm{d}}x}
\end{eqnarray}
\end{subequations}
one obtains integrals that are not convolutions. Choosing as Fourier pair $p(x)$ and $\varphi(u)$, defined in eq.(\ref{equ1}) and letting $u=0$, it is easy to obtain the generalized form of eq.(\ref{equ3}) from the latter, that can be written as
\begin{subequations}\label{equ9}
\begin{eqnarray}\label{equ9a}
\left( {I_ \pm ^\gamma  \varphi } \right)\left( 0 \right) = \int_{ - \infty }^\infty  {\left( { \pm {\rm{i}}x} \right)^{ - \gamma } p\left( x \right){\rm{d}}x}  = \left\langle {\left( { \pm {\rm{i}}X} \right)^{ - \gamma } } \right\rangle  
\end{eqnarray}
\begin{eqnarray}\label{equ9b}
 \left( {{D}_ \pm ^\gamma  \varphi } \right)\left( 0 \right)=\left( {{\bm D}_ \pm ^\gamma  \varphi } \right)\left( 0 \right) = \int_{ - \infty }^\infty  {\left( { \pm {\rm{i}}x} \right)^\gamma  p\left( x \right){\rm{d}}x = } \left\langle {\left( { \pm {\rm{i}}X} \right)^\gamma  } \right\rangle 
\end{eqnarray}
\end{subequations}

The set of complex moments is a natural generalization of integer moments as like as RL fractional differential operators generalize the classical differential calculus and, coherently eq.(\ref{equ9b}) coincides with eq.(\ref{equ3}), when $\gamma$ assumes integer values. The latter equations relate the behavior of the characteristic function and the existence of moments in a more direct way than those proposed up to now in literature \cite{wolf75}. Moreover, in contrast to the local nature of the derivative and consequently of the integer moments, the complex moments are non-local as like the fractional derivatives.

Whilst eqs.(\ref{equ9a}) and (\ref{equ9b}) descend from a proper application of the Fourier properties of the RL operators, the generalization of eq.(\ref{equ2}) is not trivial; indeed, it relies on a generalized form of Taylor series. In \cite{cott09} various generalizations of the Taylor expansion presented in literature have been analyzed, showing that the integral Taylor form based on inverse Mellin transform, sketched in the book of Samko et al. (\cite{samk93}, pp.144-5), is the most useful to our scope. The starting point to find such a generalized Taylor form is to interpret the fractional derivative and integral defined in eq.(\ref{equ5}) as a Mellin transform. In what follows we develop our proof just for a characteristic function $\varphi (s)$ for the sake of simplicity, but the result obtained here is more general. Moreover, from this point on, we will use only the definition of the Marchaud  fractional derivative because same concepts follow plainly for the RL definition. Then, letting $x=0$, eqs.(\ref{equ5}) can be written as
\begin{subequations}\label{equ10}
\begin{eqnarray}\label{equ10a}
\Gamma \left( \gamma  \right)\left( {I_ \pm ^\gamma  \varphi } \right)\left( 0 \right) = {\cal M}\left\{ {\varphi \left( { \mp \xi } \right),\gamma } \right\}
\end{eqnarray}
\begin{eqnarray}\label{equ10b}
\Gamma \left( { - \gamma } \right)\left( {{\bm D}_ \pm ^\gamma  \varphi } \right)\left( 0 \right) = {\cal M}\left\{ {\varphi \left( { \mp \xi } \right) - \varphi \left( 0 \right), - \gamma } \right\} 
\end{eqnarray}
\end{subequations}
where $\cal{M\{\cdot,\gamma\}}$ is the Mellin transform. Applying the inverse Mellin transform and recalling the condition $\varphi (0)=1$, two representations of the characteristic function 
\begin{subequations}\label{equ11}
\begin{eqnarray}\label{equ11a}
\varphi \left( { \mp u} \right) = \frac{1}{{2\pi {\rm{i}}}}\int_{\rho  - {\rm{i}}\infty }^{\rho  + {\rm{i}}\infty } {\Gamma \left( \gamma  \right)\left( {I_ \pm ^\gamma  \varphi } \right)\left( 0 \right)u^{ - \gamma } {\rm{d}}\gamma } 
\end{eqnarray}
\begin{eqnarray}\label{equ11b}
 \varphi \left( { \mp u} \right) = 1 + \frac{1}{{2\pi {\rm{i}}}}\int_{\rho  - {\rm{i}}\infty }^{\rho  + {\rm{i}}\infty } {\Gamma \left( { - \gamma } \right)\left( {{\bm D}_ \pm ^\gamma  \varphi } \right)\left( 0 \right)u^\gamma  {\rm{d}}\gamma }
\end{eqnarray}
\end{subequations}
hold true, with $u>0$ and where integrals are performed along the imaginary axis with fixed real $\rho $ that belongs to the so called fundamental strip of the Mellin transform of the function $\varphi \left( u \right)$.

The latter equations represent integral forms of the Taylor expansion for the CF, in the sense that from knowledge of the function $\left( {I_ \pm ^\gamma  \varphi } \right)\left( 0 \right)$ or $\left( {\bm{D}_ \pm ^\gamma  \varphi } \right)\left( 0 \right)$ it is possible to evaluate the function $\varphi \left( u \right)$ by integration along the imaginary line with real part $\rho $ (\cite{samk93}, pp.144-5). 

Such a representation gains an appealing touch once eqs.(\ref{equ9}) are considered, leading to
\begin{subequations}\label{equ12}
\begin{eqnarray}\label{equ12a}
 \varphi \left( { \mp u} \right) = \frac{1}{{2\pi {\rm{i}}}}\int_{\rho  - {\rm{i}}\infty }^{\rho  + {\rm{i}}\infty } {\Gamma \left( \gamma  \right)\left\langle {\left( { \pm {\rm{i}}X} \right)^{ - \gamma } } \right\rangle u^{ - \gamma } {\rm{d}}\gamma } 
\end{eqnarray}
\begin{eqnarray}\label{equ12b}
 \varphi \left( { \mp u} \right) = 1 + \frac{1}{{2\pi {\rm{i}}}}\int_{\rho  - {\rm{i}}\infty }^{\rho  + {\rm{i}}\infty } {\Gamma \left( { - \gamma } \right)\left\langle {\left( { \pm {\rm{i}}X} \right)^\gamma  } \right\rangle u^\gamma  {\rm{d}}\gamma }
\end{eqnarray}
\end{subequations}
with $u>0$. The latter equations are the integral extensions of eq.(\ref{equ2}) searched. It must be noted that the presence of the Gamma function in the inverse Mellin transform turns into a very amenable numerical treatment, as pointed out in the following.
\\

The quite simple way we derived the latter relations entails many aspects about analytic functions, due to the complex nature of the inverse Mellin transform involved. A few remarks are therefore in order: $i)$ Passing from eq.(\ref{equ12a}) to eq.(\ref{equ12b}) it does not suffice a change of sign because of the presence of the integrand residue in $\gamma=0$. $ii)$ It is not difficult to show that the integrand of eq.(\ref{equ12a}) might have, at most, poles for $\varphi(-k+\rm{i} 0)$, $k=0,1,2...$ and, conversely, the integrand in eq.(\ref{equ12b})  might have, at most, poles for $\varphi(k+\rm{i} 0)$, $k=1,2...$. $iii)$ The integral in eq.(\ref{equ12}) coincides with the sum of all the residues, and such sum can be proved to correspond to the r.h.s. of eq.(\ref{equ2}). Distributions with divergent moments can therefore be represented as sum of the residues of the integrand eq.(\ref{equ12}). Such an interesting theoretical aspect cannot be derived from classical applications of the inverse Mellin transform in probability, because in such works the starting point is the density and not the characteristic function.
\\

Of course, the fundamental strip depends on the integrability of the CF,following eqs(\ref{equ10}). It has been proved in \cite{cott09} that the strictest fundamental strip associated with every absolute convergent CF, thus including  $\alpha$-stable random variables and other power-law distributions, is the interval $0 < \rho  < 1$ if one uses eq.(\ref{equ12a}). Then, in the following, we always assume such a restriction on $ \rho $. 

The density function can be restored from eq.(\ref{equ12a}) by inverse Fourier transform, that explicitly reads
\begin{eqnarray}
  p(x)= {\cal F}^{ - 1} \left\{ {\varphi \left( u \right);x} \right\} = \frac{1}{{\left( {2\pi } \right)^2 \,{\rm{i}}}}\int_{\rho  - {\rm{i}}\infty }^{\rho  + {\rm{i}}\infty } {\Gamma \left( \gamma  \right)\left\{ {\left( {I_ + ^\gamma  \varphi } \right)\left( 0 \right)\int_0^\infty  {u^{ - \gamma } e^{ - {\rm{i}}ux} {\rm{d}} u}  + } \right.}  \nonumber \\ + \left( {I_ - ^\gamma  \varphi } \right)\left( 0 \right)\left. {\int_{ - \infty }^0 {\left( { - u} \right)^{ - \gamma } e^{ - {\rm{i}}ux} {\rm{d}} u} } \right\}{\rm{d}}\gamma
\label{equ13}
\end{eqnarray}
Further simplification leads to
\begin{eqnarray}
  p(x)= \frac{1}{{\left( {2\pi } \right)^2 }}\int_{\rho  - i\infty }^{\rho  + i\infty } {\Gamma \left( \gamma  \right)\Gamma \left( {1 - \gamma } \right)\left\{ {\langle {\left( { - {\rm{i}}X} \right)^{ - \gamma } } \rangle\left( {{\rm{i}}x} \right)^{\gamma  - 1} } \right.}  +  \left. {  \langle {\left( {\rm{i}} X \right)^{ - \gamma } } \rangle\left( { - {\rm{i}} x} \right)^{\gamma  - 1} } \right\}{{\rm{d}}}\gamma 
\label{equ14}
\end{eqnarray}

Eq.(\ref{equ14}) represents the density function by the integral form of the Taylor approximation and can be interpreted as the generalization of eq.(\ref{equ4}). We note that the density is expressed in integral form without the need of partitioning, and no restriction on the non-negativity of the density is needed. This fact is the consequence of having used the Mellin transform of the characteristic function. A further simplification is possible in virtue of the Hermitian property of the characteristic function, since $\varphi \left( u \right) = \overline{\varphi (-u)}$ where the overbar means conjugation. Evaluation of the
CF for $u > 0$ suffices to restore the PDF in the form
\begin{equation}
p\left( x \right) = \frac{2}{{\left( {2\pi } \right)^2 }}{\rm{Re}} \left\{ {\int_{\rho  - \rm{i}\infty }^{\rho  + \rm{i}\infty } {\frac{\pi }{{\sin \left( {\pi \gamma } \right)}}\langle {\left( { - {\rm{i}} X} \right)^{ - \gamma } } \rangle\left( {\rm{i}} x \right)^{\gamma  - 1} } {\rm{d}} \gamma } \right\}\,
\label{equ15}
\end{equation}
where the property $\pi /\sin \left( {\pi \gamma } \right) = \Gamma \left( {1 - \gamma } \right)\Gamma \left( \gamma  \right)$ has been used. By looking at eq.(\ref{equ15}) one may state that the PDF in terms of complex moments remains meaningful and computationally useful. Moreover, since the integration is performed only on the imaginary axis, i.e. the real part of $\gamma$ remains fixed in the integrand, $\langle {\left( { - {\rm{i}}X} \right)^{ - \gamma } } \rangle$ exists also for distributions for which $\langle {X^j } \rangle$ does not exist, for $j$ integer and greater than a certain value. By comparing eqs.(\ref{equ12}) and (\ref{equ15}) a perfect duality between the representation of the PDF and the CF is evidenced. The integrals in eqs.(\ref{equ12a}) and (\ref{equ15}) remain independent of the choice of $\rho$ within the strip $0 < \rho  < 1$, because of the holomorphy of the integrand in the fundamental strip.

The appeal of such a representation appears evident when the discretization of the integral is performed. Indeed, it will be shown in the numerical section that the integral performed on the imaginary axis may be truncated at a certain value $\bar{ \eta} $, because the integrand vanishes rapidly, and therefore the discretization of such an integral always produces very accurate results, in particular for distributions with heavy tails.

\section{Extension to Multivariate Random Vectors}
In this section, the concepts briefly outlined in the previous section will be derived in case of multivariate random vectors. Multivariate random vectors are fully characterized in probabilistic setting by the joint probability density function, or by its spectral counterpart, namely the joint characteristic function; alternatively also by joint moments or by joint cumulants of every order. Let $\left( {{\cal S},{\cal A},{\cal P}} \right)$ be a probability space, ${\bf{X}}:\left( {{\cal S},{\cal A}} \right) \to \left( {\mathbb{R} ^d ,{\cal B}^d } \right)$, that is ${\bf{X}}^T  = \left( {X_1 ,...,X_d } \right)$ denotes a d-dimensional random vector. As usual, by denoting ${\bf{x}}^T  = \left( {x_1 ,...,x_d } \right)$, the joint distribution function of $\bf{X}$ is defined as $F\left( {\bf{x}} \right) = \Pr \left( {X_1  < x_1 ,....,X_d  < x_d } \right)$ and its derivative is called joint probability density function, namely $p\left( {\bf{x}} \right) = \partial ^d F\left( {\bf{x}} \right)/\partial x_1 ...\partial x_d $. Let the vector ${\bf{u}}^T = \left( {u_1 ,...,u_d } \right)$, with $u_j  \in \mathbb{R} $; the joint characteristic function $\varphi \left( {\bf{u}} \right)$ is the multidimensional Fourier transform of the joint density, that is  
\begin{equation}
\varphi \left( {\bf{u}} \right) = {\cal F}\left\{ {p\left( {\bf{x}} \right);{\bf{u}}} \right\} = \int_{ - \infty }^\infty  {...\int_{ - \infty }^\infty  {\exp \left( {{\rm{i}}{\bf{u}}^T {\bf{x}}} \right)\,p\left( {\bf{x}} \right){\rm{d}} x_1 ...{\rm{d}} x_d } } 
\label{equ16}
\end{equation}
 
Analogously to the one dimensional case, we first define joint complex moments 
\begin{equation}
  \langle {\left( { \mp {\rm{i}} X_1 } \right)^{ - \gamma _1 } ...\,\,\left( { \mp {\rm{i}} X_d } \right)^{ - \gamma _d } } \rangle = \int_{ - \infty }^\infty  {...\int_{ - \infty }^\infty  {\left( { \mp {{\rm{i}}} x_1 } \right)^{ - \gamma _1 } ... \left( { \mp {\rm{i}} x_d } \right)^{ - \gamma _d } p\left( {\bf{x}} \right){\rm{d}} x_1 ...{\rm{d}} x_d } }
\label{equ17}
\end{equation}

It will be shown that they are multidimensional fractional derivatives of the joint characteristic function evaluated in zero. To this purpose, let ${\bf{x}}^T {\rm{ = }}\left( {x_{\rm{1}} ,...,x_d } \right)$, ${\bm{\gamma} }^T =\left( {\gamma _{\rm{1}} ,...,\gamma _d } \right)$ be elements of $\mathbb{R}^d $ and $f\left( {\bf{x}} \right)$  a function of $d$ variables. As reported in (\cite{samk93}, p. 462), the Liouville-type fractional integral of order $\gamma$ with ${\rm{ Re}} \gamma _k  > 0,\,k = 1,...,d$, is defined as
\begin{equation}
  (I_{ \pm ... \pm }^{\bm{\gamma} } f )\left( {\bf{x}} \right) = \frac{1}{{\Gamma \left( {\gamma _1 } \right)...\Gamma \left( {\gamma _d } \right)}}\int_0^\infty  {...\int_0^\infty  {\xi _1^{\gamma _1  - 1} ...\xi _d^{\gamma _d  - 1} f\left( {x_1  \mp \xi _1 ,...,x_d  \mp \xi _d } \right){{\rm{d}}}\xi _1 ...{{\rm{d}}}\xi _d } } 
\label{equ18}
\end{equation}
where the choice of signs must be coherent between the left and the right hand sides. Let the function $f\left( {\bf{x}} \right)$ be defined on the whole space $\mathbb{R}^d$ such that $\int\limits_{\mathbb{R}^d } {\left| {f\left( \bf{x} \right)} \right|{{\rm{d}}}x_1 ...{{\rm{d}}}x_d {\rm{ <  }}\infty }$ and indicate the multidimensional Fourier transform of $f\left( {\bf{x}} \right)$ as usual in the form 
\begin{equation}
\left( {{\cal F}f} \right)\left( \bf{u}  \right) = {\cal F}\left\{ {f\left( {\bf{x}} \right);\bf{u}} \right\}\mathop  = \limits^{def} \int\limits_{\mathbb{R} ^d } {e^{{\rm{i}}\left( {u _1 \,\xi _1  + ... + u _d \,\xi _d } \right)} f\left( \bm{\xi}  \right){\rm{d}}\xi _{\rm{1}} ...{\rm{d}}\xi _d } 
\label{equ19}
\end{equation}
and the multidimensional Mellin transform as
\begin{equation}
\left( {{\cal M}f} \right)\left( \bm{\gamma}  \right) = {\cal M}\left\{ {f\left( {\bf{x}} \right);\bm{\gamma} } \right\}\mathop  = \limits^{def} \int_0^\infty  {...\int_0^\infty  {\xi _{\rm{1}}^{\gamma _{\rm{1}}  - 1} ...\,\xi _d^{\gamma _d  - 1} f\left( \bm{\xi} \right){\rm{d}}\xi _1 ...{\rm{d}}\xi _d } } 
\label{equ20}
\end{equation}

The multidimensional Fourier transform of the Liouville-type fractional integral of $f\left( {\bf{x}} \right)$ is related to the Fourier transform of $f\left( {\bf{x}} \right)$ by the following expressions 
\begin{subequations}\label{equ21}
\begin{eqnarray}
{\cal F}\left\{ {\left( {I_{ \pm ... \pm }^{\bm{\gamma} } f} \right)\left( {\bf{x}} \right);\,{\bf{u} }} \right\} = \left( { \mp {\rm{i}} u_1 } \right)^{ - \gamma _1 } ...\,\,\left( { \mp {\rm{i}} u _d } \right)^{ - \gamma _d } \left( {{\cal F}f} \right)\left( {\bf{u} } \right)\label{equ21a}
\end{eqnarray}
\begin{eqnarray}
{\cal F}^{ - {1}} \left\{ {\left( {I_{ \pm ... \pm }^{\bm{\gamma}} f} \right)\left( {\bf{x}} \right);\,\bf{u}} \right\} = \left( { \pm {\rm{i}} u _1 } \right)^{ - \gamma _1 } ...\,\,\left( { \pm {\rm{i}} u _d } \right)^{ - \gamma _d } \left( {{\cal F}^{ - {\rm{1}}} f} \right) \left( \bf{u} \right)\label{equ21b}
\end{eqnarray}
\end{subequations}
as reported in (\cite{samk93}, p. 474). Now, we show that joint complex moments arise in a natural way once the fractional integral of the joint characteristic function is calculated in zero. Indeed, eq.(\ref{equ21b}) specified for the joint characteristic function gives the relation
\begin{equation}
{\cal F}^{-1}\left\{ {\left( {I_{ \pm ... \pm }^{{\bm{\gamma} }} \varphi } \right)\left( {\bf{u}} \right);\,{\bf{x}}} \right\} = \left( { \pm {\rm{i}} x_1 } \right)^{ - \gamma _1 } ...\,\,\left( { \pm {\rm{i}} x_d } \right)^{ - \gamma _d } {\cal F}^{-1}\left\{ {\varphi \left( {\bf{u}} \right);{\bf{x}}} \right\}
\label{equ22}
\end{equation}

The multidimensional inverse Fourier transform of the joint characteristic function is the joint PDF and therefore, a multidimensional Fourier transform yields
\begin{equation}
\left( {I_{ \pm ... \pm }^{{\bm{\gamma} }} \varphi } \right)\left( {\bf{u}} \right) = \int_{ - \infty }^\infty  {...\int_{ - \infty }^\infty  {\left( { \pm {\rm{i}} x_1 } \right)^{ - \gamma _1 } ...\left( { \pm {\rm{i}} x_d } \right)^{ - \gamma _d } p\left( {\bf{x}} \right)\exp \left( {{{\rm{i}}}{\bf{u}}^T {\bf{x}}} \right){\rm{d}} x_1 ...{\rm{d}} x_d } } 
\label{equ23}
\end{equation}
that, calculated in ${\bf{u}}^T  = \left( {0,...,0} \right)$ gives the fundamental relation 
\begin{equation}
I_{ \pm ... \pm }^{{\bm{\gamma} }} \varphi \left( {{\rm{0}},...,{\rm{0}}} \right) = \langle {\left( { \pm {{\rm{i}}}X_1 } \right)^{ - \gamma _1 } ...\left( { \pm {{\rm{i}}}X_d } \right)^{ - \gamma _d } } \rangle
\label{equ24}
\end{equation}

With the same reasoning, it is possible to relate the fractional derivatives of the joint characteristic function calculated in zero in terms of complex moments as
\begin{equation}
{\bm D}_{ \pm ... \pm }^{{\bm{\gamma} }} \varphi \left( {{\rm{0}},...,{\rm{0}}} \right) = \langle {\left( { \pm {{\rm{i}}}X_1 } \right)^{\gamma _1 } ...\left( { \pm {{\rm{i}}}X_d } \right)^{\gamma _d } } \rangle
\label{equ25}
\end{equation}

Once these results are achieved, we derive a generalized integral form of the Taylor series. Such an expansion may be obtained by applying the commutative property of integro-differential operators and by the inverse Mellin transform also in the multidimensional case. To this aim, we interpret the Liouville-type multidimensional fractional integral as the multidimensional Mellin transform of the function $f\left( {\bf{x}} \right)$
\begin{eqnarray}
  \Gamma \left( {\gamma _1 } \right)...\Gamma \left( {\gamma _d } \right)\left( {I_{ \pm ... \pm }^{{\bm{\gamma} }} f} \right)\left( {x_1 ,...,x_d } \right) = \int\limits_0^\infty  {...} \int\limits_0^\infty  {\xi _1^{\gamma _1  - 1} ...\xi _d^{\gamma _d  - 1} }  \times \nonumber \\ 
 \times f\left( {x_1  \mp \xi _1 ,...,x_d  \mp \xi _d } \right){\rm{d}}\xi _1 ...{\rm{d}}\xi _d  ,
\label{equ26}
\end{eqnarray}
evaluated at fixed ${\bf{x}}^T  = \left( {x_1 ,...,x_d } \right)$. Writing eq.(\ref{equ26}) for the joint characteristic function, evaluated in the point ${\bf{u}}^T  = \left( {0,...,0} \right)$
\begin{eqnarray}
  \Gamma \left( {\gamma _1 } \right)...\Gamma \left( {\gamma _d } \right)\left( {I_{ \pm ... \pm }^{{\bm{\gamma} }} \varphi } \right)\left( {0,...,0} \right) = \int\limits_0^\infty  {...} \int\limits_0^\infty  {u_1^{\gamma _1  - 1} ...u_d^{\gamma _d  - 1} } \times \nonumber \\ 
\times \varphi \left( { \mp u_1 ,..., \mp u_d } \right){\rm{d}} u_1 ...{\rm{d}} u_d ,
\label{equ27}
\end{eqnarray}
it is easy to recognize that the r.h.s is the Mellin transform, eq.(\ref{equ20}), of the l.h.s.. Introducing eq.(\ref{equ24}), the former assumes the relevant form
\begin{eqnarray}
   \Gamma \left( {\gamma _1 } \right)...\Gamma \left( {\gamma _d } \right)\langle {\left( { \mp {\rm{i}} X_1 } \right)^{ - \gamma _1 } ...\left( { \mp {\rm{i}} X_d } \right)^{ - \gamma _d } } \rangle\, = \int\limits_0^\infty  {...} \int\limits_0^\infty  {u_1^{\gamma _1  - 1} ...u_d^{\gamma _d  - 1} }  \times  \nonumber \\ 
\times \,\varphi \left( { \mp u_1 ,..., \mp u_d } \right){\rm{d}} u_1 ...{\rm{d}} u_d .
\label{equ28}
\end{eqnarray}

By performing the inverse Mellin transform, the joint characteristic function is rewritten as 
%	\varphi \left( {u_1 ,...,u_d } \right) = \frac{1}{{\left( {2\pi {{\rm{i}}}} \right)^d }}\int\limits_{\rho _1  - {{\rm{i}}}\infty }^{\rho _1  + {{\rm{i}}}\infty } {...} \int\limits_{\rho _d  - {{\rm{i}}}\infty }^{\rho _d  + {{\rm{i}}}\infty } {\Gamma \left( {\gamma _1 } \right)...\Gamma \left( {\gamma _d } \right)}  \times  \nonumber  \\ 
%\times \, \langle {\left( { \mp _1 {{\rm{i}}}X_1 } \right)^{ - \gamma _1 } ...\left( { \mp _d {{\rm{i}}}X_d } \right)^{ - \gamma _d } } \rangle\left| {u_1 } \right|^{ - \gamma _1 } ...\left| {u_d } \right|^{ - \gamma _d } {\rm{d}}\gamma _1 ...{\rm{d}}\gamma _d 
\begin{eqnarray}
	\varphi \left( {u_1 ,...,u_d } \right) = \frac{1}{{\left( {2\pi {{\rm{i}}}} \right)^d }}\int\limits_{\rho _1  - {{\rm{i}}}\infty }^{\rho _1  + {{\rm{i}}}\infty } {...} \int\limits_{\rho _d  - {{\rm{i}}}\infty }^{\rho _d  + {{\rm{i}}}\infty } {\Gamma \left( {\gamma _1 } \right)...\Gamma \left( {\gamma _d } \right)}  \times  \nonumber  \\ 
\times \, \langle {\left( { - {\rm{sign}}( {u_1 }) \; {{\rm{i}}}X_1 } \right)^{ - \gamma _1 } ...\left( { - {\rm{sign}}( {u_d }) \; {{\rm{i}}}X_d } \right)^{ - \gamma _d } } \rangle \times  \nonumber  \\ 
\times \left| {u_1 } \right|^{ - \gamma _1 } ...\left| {u_d } \right|^{ - \gamma _d } {\rm{d}}\gamma _1 ...{\rm{d}}\gamma _d 
\label{equ29}
\end{eqnarray}

The integrals are performed by choosing $\rho _1 ,...,\rho _d $ inside the fundamental multidimensional strip. Following \cite{sriv95}, assume that $\varphi \left( {u_1 ,...,u_d } \right)$ is such that 
\begin{equation}
  \varphi \left( {u_1 ,...,u_d } \right) = O\left( {\left| {u_1 } \right|^{ - q_1 } ,...,\left| {u_d } \right|^{ - q_d } } \right)\,\,\,\,\,\,for\,\,\,\min \left( {\left| {u_1 } \right|,...,\left| {u_d } \right|} \right) \to \infty 
\label{equ30}
\end{equation}
with $q_j  > 0$, $j = 1,...,d$. Then, the multidimensional Mellin transform in eq.(\ref{equ20}) is meaningful under the conditions $0 < \rho _j  < q_j $. Due to the absolute convergence of the joint characteristic function of a true multivariate random variable (see \cite{cott09}) one can select the strictest fundamental strip $0 < \rho _j  < 1$ in which eq.(\ref{equ29}) holds true. 

As in the one dimensional case, from the knowledge of the joint characteristic function one can restore the joint PDF by Fourier inversion. The multidimensional inverse Fourier transform of relation (\ref{equ28}) can be performed by subdivision of the domain $\mathbb{R} ^d $ in $m = 2^d $ subsets in which the variables ${\bm{u}}^T  = \left( {u_1 ,...,u_d } \right)$ assume all the possible combinations of positive and negative values. By means of some algebra, the multidimensional inverse Fourier transform of eq.(\ref{equ29}) leads to the expression of the joint PDF of the random vector $\bm{X}$ , that reads
\begin{eqnarray}
   p\left( {x_1 ,...,x_d } \right) = \frac{1}{{\left( {2\pi i} \right)^d \left( {2\pi } \right)^d }}\int\limits_{\rho _1  - {{\rm{i}}}\infty }^{\rho _1  + {{\rm{i}}}\infty } {...} \int\limits_{\rho _d  - {{\rm{i}}}\infty }^{\rho _d  + {{\rm{i}}}\infty } {\frac{\pi }{{\sin \left( {\pi \gamma _1 } \right)}}...\frac{\pi }{{\sin \left( {\pi \gamma _d } \right)}} \times } \nonumber  \\ 
  \times \sum\limits_{k_1  = 1}^2 {...\sum\limits_{k_d  = 1}^2 {\langle {\left( {\left( { - 1} \right)^{k_1 } {{\rm{i}}}X_1 } \right)^{ - \gamma _1 } ...\left( {\left( { - 1} \right)^{k_{\rm{d}}  } {{\rm{i}}}X_{\rm{d}}  } \right)^{ - \gamma _{\rm{d}}  } } \rangle} }  \times  \nonumber \\ 
\times \left( {\left( { - 1} \right)^{k_1  + 1} {{\rm{i}}}\,x_1 } \right)^{\gamma _1  - 1} ...\left( {\left( { - 1} \right)^{k_{\rm{d}}   + 1} {{\rm{i}}}\,x_d } \right)^{\gamma _d  - 1} {\rm{d}}\gamma _1 ...{\rm{d}}\gamma _d ,
\label{equ31}
\end{eqnarray}
where the relation $\pi /\sin \left( {\pi \gamma _j } \right) = \Gamma \left( {1 - \gamma _j } \right)\Gamma \left( {\gamma _j } \right)$ has been used. Eq.(\ref{equ31}) can be simplified recalling that the joint characteristic function has symmetry properties. Indeed, $\varphi \left( {\bf{u}} \right) =\overline{\varphi \left( { - {\bf{u}}} \right)}$ and the joint PDF can be restored by $2^{d - 1} $ subsets, in the following form
\begin{eqnarray}
  p\left( {x_1 ,...,x_d } \right) = 2{\mathop{\rm Re}\nolimits} \left\{ {\frac{1}{{\left( {2\pi {{\rm{i}}}} \right)^d \left( {2\pi } \right)^d }}\int\limits_{\rho _1  - {{\rm{i}}}\infty }^{\rho _1  + {{\rm{i}}}\infty } {...} \int\limits_{\rho _d  - {{\rm{i}}}\infty }^{\rho _d  + {{\rm{i}}}\infty } {\frac{\pi }{{\sin \left( {\pi \gamma _1 } \right)}}...\frac{\pi }{{\sin \left( {\pi \gamma _d } \right)}} \times } } \right. \nonumber \\ 
  \times \sum\limits_{k_2  = 1}^2 {...\sum\limits_{k_d  = 1}^2 {\langle {\left( { - {{\rm{i}}}X_1 } \right)^{ - \gamma _1 } \left( {\left( { - 1} \right)^{k_2 } {{\rm{i}}}X_2 } \right)^{ - \gamma _2 } ...\left( {\left( { - 1} \right)^{k_d } {{\rm{i}}}X_d } \right)^{ - \gamma _d } } \rangle} }  \times \nonumber  \\ 
 \left. { \times \left( {{{\rm{i}}}x_1 } \right)^{\gamma _1  - 1} \left( {\left( { - 1} \right)^{k_2  + 1} {{\rm{i}}}x_2 } \right)^{\gamma _2  - 1} ...\left( {\left( { - 1} \right)^{k_d  + 1} {{\rm{i}}}x_d } \right)^{\gamma _d  - 1} {\rm{d}}\gamma _1 ...{\rm{d}}\gamma _d } \right\} \label{equ32}
\end{eqnarray}

We note that in eq.(\ref{equ32}) there occur $d - 1$ summations and the first variable is not indexed. Concluding, from eq.(\ref{equ29}) and (\ref{equ32}) we can affirm that the function  
\[
\langle {\left( { - {\rm{i}}X_1 } \right)^{ - \gamma _1 } \left( {\left( { - 1} \right)^{k_2 } {\rm{i}}X_2 } \right)^{ - \gamma _2 } ...\left( {\left( { - 1} \right)^{k_d } {\rm{i}}X_d } \right)^{ - \gamma _d } } \rangle
\]
is able to represent both the joint PDF and the joint characteristic function. In the following section, it will be shown that discretization produces the desired results. That is, with a finite number of fractional moments, both the joint PDF and the joint characteristic function can be restored.

\section{Applications}
In this section it will be shown how previous results can be used to characterize probability distributions. The integral forms in the monovariate case, eqs.(\ref{equ12}) and (\ref{equ14}), and in the multivariate case, eqs.(\ref{equ29}) and (\ref{equ31}), will be discretized in a suitable form. A remarkable effect of having applied the Mellin transform to the characteristic function rather than to the density is the presence of the gamma function in the integrand. Indeed, by its fast decay with increasing imaginary part of the argument, it makes the integral amenable to numerical calculations.

\subsection{Numerical treatment of the integral representation by complex moments for random variables}
The integrals in eqs.(\ref{equ12}) will be evaluated considering a partition of the truncated domain $[-\bar\eta, \bar\eta]$ in $2m$ ($m\in \mathbb{N}$) intervals of amplitude $\Delta \eta $, such that $\bar\eta=m \Delta \eta$. In this way the integrals are approximated as a sum evaluated at ${\gamma _{k} }=\rho+ {{\rm{i}}} k \Delta\eta$:
\begin{subequations}\label{equ33}
\begin{eqnarray}
\varphi \left( u \right) = \frac{{\Delta \eta }}
{{2\pi }}\sum\limits_{k = 0}^{2m + 1} {\Gamma \left( { \gamma _{k - m} } \right)\left\langle {\left( { -{{\rm{i}}} X} \right)^{-\gamma _{k  - m}} } \right\rangle u^{-\gamma _{k - m} } } 
\end{eqnarray}\label{equ33a}
\begin{eqnarray}
\varphi \left(  u \right) = 1 + \frac{{\Delta \eta }}
{{2\pi }}\sum\limits_{k = 0}^{2m + 1} {\Gamma \left( { - \gamma _{k - m} } \right)\left\langle {\left( { - {{\rm{i}}} X} \right)^{\gamma _{k  - m}} } \right\rangle u^{\gamma _{k - m} } } 
\end{eqnarray}\label{equ33b}
\end{subequations}
and are valid for $u>0$. For $u<0$ the characteristic function is restored by the property $\varphi \left( u \right) = \overline{\varphi (-u)}$.
As mentioned, the truncation is feasible because of the presence of the gamma function that vanishes as the imaginary variable $\eta$ increases. Consequently, according to eq.(\ref{equ15}), the density can be evaluated by means of the same complex moments in the form
\begin{equation}
p\left( x \right) = \frac{{\Delta \eta }}
{{2\pi }}{\rm{Re}} \left[ {\sum\limits_{k = 0}^{2m + 1} {\left\langle {\left( { - iX} \right)^{ - \gamma _{k - m} } } \right\rangle \frac{{\left( {ix} \right)^{\gamma _{k - m}  - 1} }}
{{{\rm{sin}}\left( {\pi \gamma _{k - m} } \right)}}} } \right]
\label{equ34}
\end{equation}
and, then, the calculation of $2m +1$ complex moments suffices to have a complete probabilistic description of the random variable.

\subsubsection{Example: Power-law tail distributions}
\begin{figure}[b]
\includegraphics[width=\columnwidth]{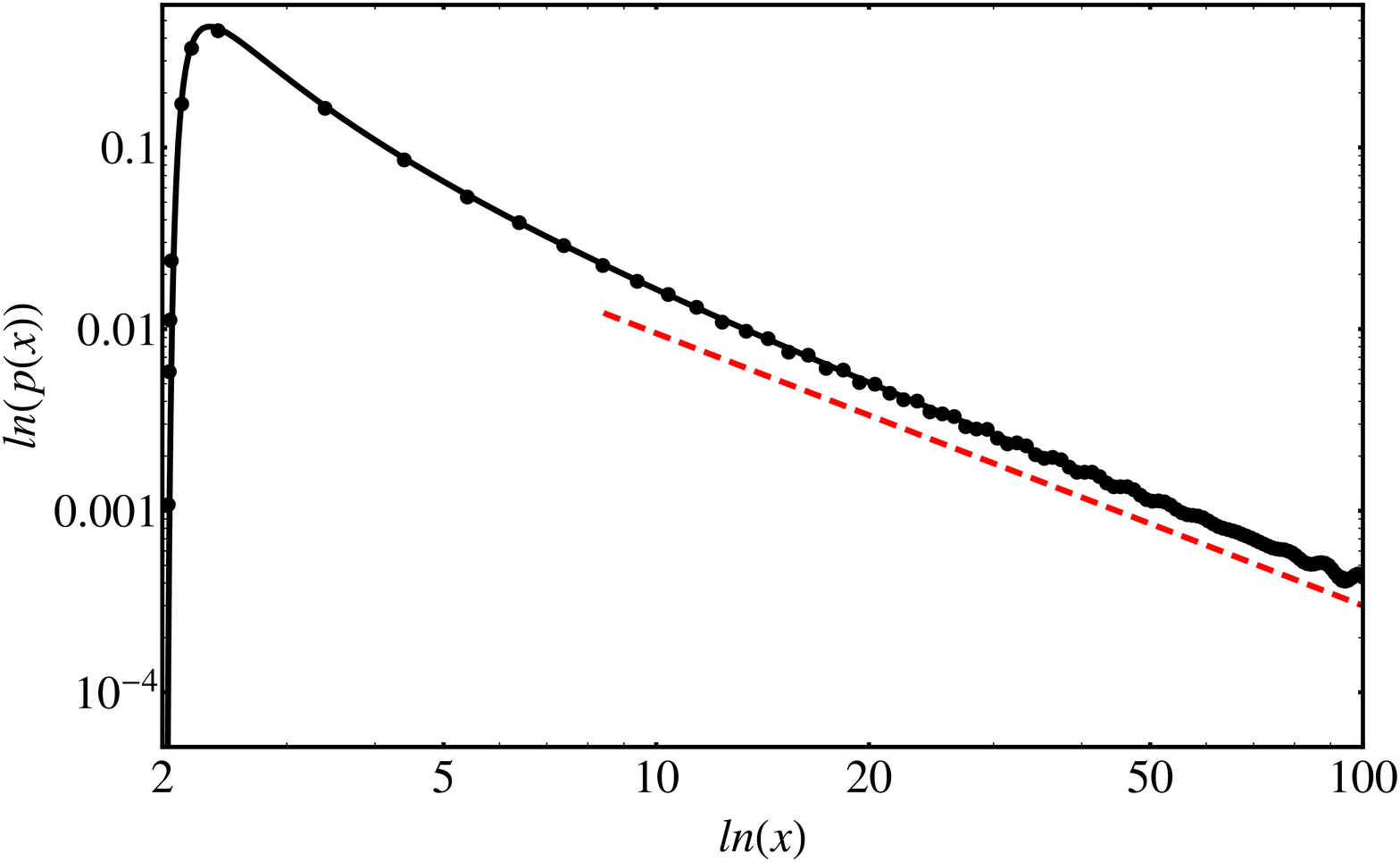}
\caption{Comparison between the exact (continuous) and the approximated (dotted) density of a 1/2-stable distribution with $\sigma =1.0$, $\mu=2$; power-law trend is plotted in dashed line.}
\label{figlevy}
\end{figure}

\begin{figure}[b]
\includegraphics[width=\columnwidth]{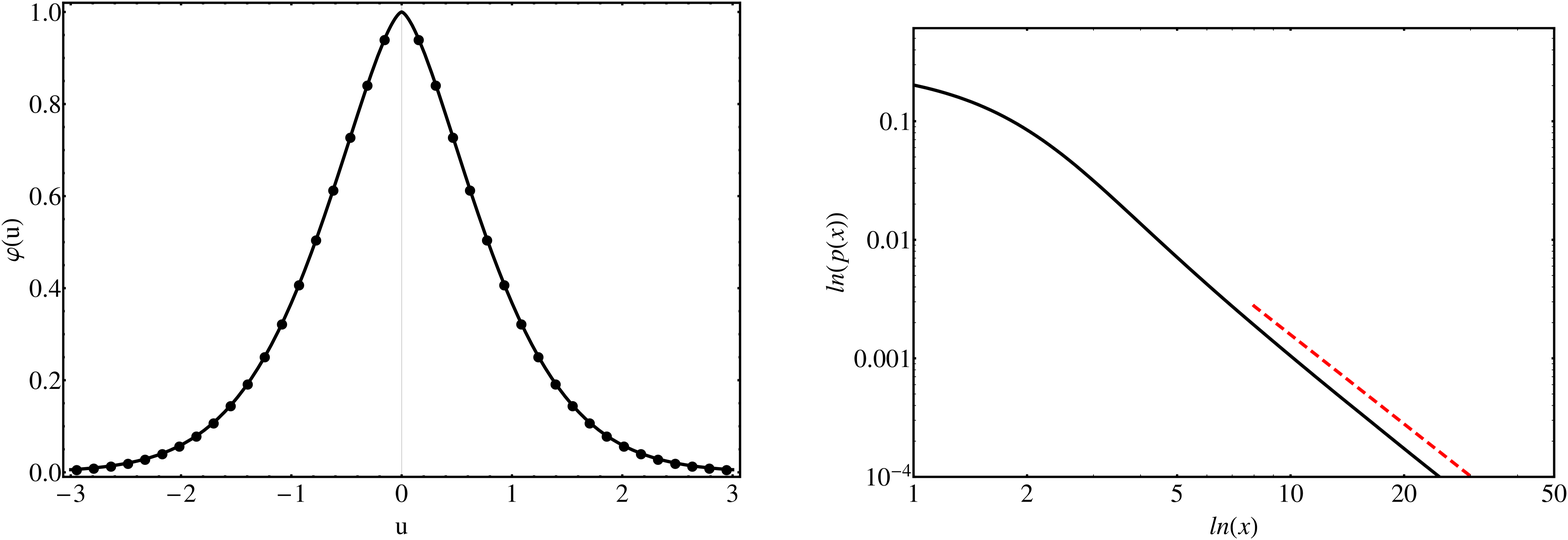}
\caption{(Left panel) comparison between the exact (continuous) and the approximated (dotted) characteristic function of a symmetric 3/2-stable distribution; (right panel) corresponding density in log-log diagram, contrasted with power-law trend (dashed line).}
\label{figstable15}
\end{figure}

Firstly, we consider an $\alpha$-stable random variable, with $\alpha =1/2$, whose characteristic function is known to be of the form \cite{samo94}
 \[
\varphi \left( u \right) = e^{ - \sigma ^{1/2} \left| u \right|^{1/2} \left( {1 - {\rm i} \; {\rm{sign}}\left( {\rm{u}} \right)} \right) + {\rm i}\mu u} 
\]
where $\sigma$ and $\mu$ are the scale and the location parameters, respectively. This distribution, also known as L\'evy-Smirnov distribution, has moments $\left\langle {X^p } \right\rangle  < \infty $ only if $p < 1/2 $. By selecting $\Delta\eta=.9$, $\rho=.85$, and $m=60$ the results plotted in the bi-logarithmic diagram in Fig.\ref{figlevy} show that in a very wide range the sum well approximates the integral. Complex moments have been calculated from eqs.(\ref{equ10})
\[
\left\langle {\left( { - {{\rm{i}}}X} \right)^{ - \gamma } } \right\rangle  = \left( {I_ - ^\gamma  \varphi } \right)\left( 0 \right) = \frac{1}
{{\Gamma \left( \gamma  \right)}}\mathcal{M}\left\{ {\varphi \left( { + u} \right);\gamma } \right\}
\]
by means of Mathematica. It has to be remarked that the results have been computed by applying eq.(\ref{equ12a}), and that $\rho=.85$ corresponds to complex moments $\left\langle {\left( {{-\rm{i}}X} \right)^{- 0.85 + i\eta }} \right\rangle$ which exist for this distribution.
As further example, let us consider the 3/2-stable distribution, known as Holtsmark distribution, whose characteristic function is 
\[
\varphi \left( u \right) = e^{ - \sigma ^{3/2} \left| u \right|^{3/2}}
\]
In this case, moments exist if $p < 3/2 $ and, in order to apply eq.(\ref{equ12a}), the value of $\rho=0.35$, i.e. moments of the type $\left\langle {\left( {{-\rm{i}}X} \right)^{- 0.35 + i\eta }} \right\rangle < \infty$ are used. Other parameters in the discretization are $\Delta\eta=.2$ and $m=50$ and the results have been reported in Fig.\ref{figstable15}. 
We stress that the complex moments for both the examples have been calculated from the knowledge of the characteristic function without explicit use of the density function. This is quite interesting with
respect to extensions of the method to random variables whose density is not known in explicit analytical form, such as $\alpha$-stable densities.

\subsubsection{Statistics from raw data}
\begin{figure}[b]
\includegraphics[width=\columnwidth]{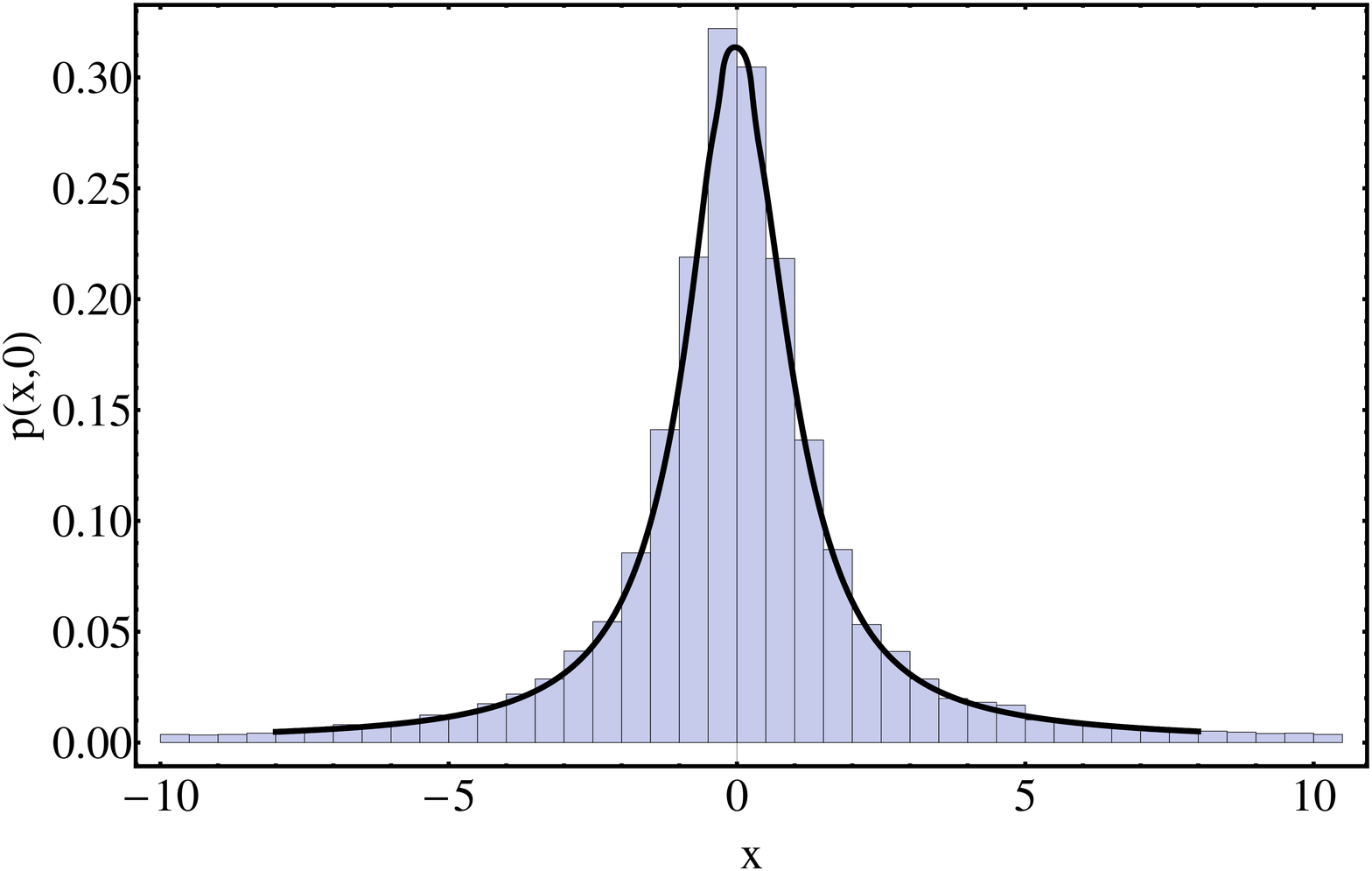}
\caption{Comparison between the density obtained by complex moments and histogram for 10000 realization of a random variable}
\label{figdata}
\end{figure}

In this example it is shown that the proposed representation is also useful when dealing with raw data from experiments.
Let us consider $N_s$ realizations of the random variable $X$ and indicate by $X_j$, $j=1,..., N_s$ the $j-th$ outcome of $X$. In order to evaluate the statistics, complex moments of order $\gamma_k = \rho + {\rm i} \eta_k $ can be calculated from the raw data via
\[
\left\langle {\left( { \pm {\rm i} X} \right)^{\rho  + {\rm i} \eta _k } } \right\rangle  = \frac{1}
{{N_s }}\sum\limits_{j = 1}^{N_s } {\left( { \pm {\rm i} X_j } \right)^{\rho  + {\rm i} \eta _k } } 
\]
for every $k=0,...,2m+1$ following the partition of the previous example. Then, applying eq.(\ref{equ34}) the density can be computed. In Fig.\ref{figdata} a comparison between the histogram and the density calculated in such a way is shown, demonstrating excellent convergence. The parameters used in this example are $N_s=10000$, $\rho=0.5$, $\Delta\eta=0.05$, $m=50$.

%\subsubsection{Statistics of the output of a non-linear stochastic differential equation}
%
%Consider the non-linear stochastic differential equation 
%\[
%\dot X\left( t \right) =  - p\left| {X\left( t \right)} \right|^{p - 1} sign\left( X \right) + W_\alpha  \left( t \right)
%\]
%with $p=1/2$ and initial condition $X(0)=N(0,1)$. 
%

\subsection{Applications to multivariate random variables}
Joint densities and characteristic functions can be represented exactly by the integral forms in eqs.(\ref{equ29}) and (\ref{equ32}) in which joint complex moments appear. In this section, the integral is computed by a simple rectangle scheme, in order to present some applications, without having to find the best numerical treatment of the integral involved. Yet, it has to be noted that more sophisticated numerical schemes (involving trapezoidal, Gaussian or other local/adaptive rules) of course may give better results. Hereinafter, the simplest integration scheme is adopted in order to show that with a finite number of quantities (the complex moments) both the joint PDF and the joint characteristic function are very well approximated. Then, indicating the integrand in eqs.(\ref{equ29}) and (\ref{equ32}) as
\[
G_\varphi  \left( {{\bf{u}},\gamma _1 ,...,\gamma _d } \right) = \frac{1}{{\left( {2\pi {\rm{i}}} \right)^d }}\Gamma \left( {\gamma _1 } \right)...\Gamma \left( {\gamma _d } \right)\left\langle {\left( { -{\rm sign}(u_1)\; {\rm{i}}X_1 } \right)^{ - \gamma _1 } ...\left( { -{\rm sign}(u_d)\; {\rm{i}}X_d } \right)^{ - \gamma _d } } \right\rangle \left| {u_1 } \right|^{ - \gamma _1 } ...\left| {u_d } \right|^{ - \gamma _d } 
\]
and
\[
\begin{array}{l}
 G_p \left( {{\bf{x}},\gamma _1 ,...,\gamma _d } \right) = \frac{1}{{\left( {2\pi i} \right)^d \left( {2\pi } \right)^d }}\frac{\pi }{{\sin \left( {\pi \gamma _1 } \right)}}...\frac{\pi }{{\sin \left( {\pi \gamma _d } \right)}} \times  \\ 
 \,\,\,\,\,\,\,\,\,\,\,\,\,\,\,\,\,\,\,\,\,\,\,\,\,\,\,\,\,\,\,\,\,\,\,\,\,\,\,\,\,\,\,\,\,\,\,\,\,\,\,\, \times \sum\limits_{k_1  = 1}^2 {...\sum\limits_{k_d  = 1}^2 {\left\langle {\left( {\left( { - 1} \right)^{k_1 } {\rm{i}}X_1 } \right)^{ - \gamma _1 } ...\left( {\left( { - 1} \right)^{k_{\mathop{\rm d}\nolimits}  } {\rm{i}}X_{\mathop{\rm d}\nolimits}  } \right)^{ - \gamma _{\mathop{\rm d}\nolimits}  } } \right\rangle  \times } }  \\ 
 \,\,\,\,\,\,\,\,\,\,\,\,\,\,\,\,\,\,\,\,\,\,\,\,\,\,\,\,\,\,\,\,\,\,\,\,\,\,\,\,\,\,\,\,\,\,\,\,\,\,\, \times \left( {\left( { - 1} \right)^{k_1  + 1} {\rm{i}}\,x_1 } \right)^{\gamma _1  - 1} ...\left( {\left( { - 1} \right)^{k_{\mathop{\rm d}\nolimits}   + 1} {\rm{i}}\,x_d } \right)^{\gamma _d  - 1}  \\ 
 \end{array}
\]
we can more compactly express eqs.(\ref{equ29}) and (\ref{equ32}) in the form
\[
\varphi \left( {\bf{u}} \right) = \int\limits_{\rho _1  - {\rm{i}}\infty }^{\rho _1  + {\rm{i}}\infty } {...} \int\limits_{\rho _d  - {\rm{i}}\infty }^{\rho _d  + {\rm{i}}\infty } {G_\varphi  \left( {{\bf{u}},\gamma _1 ,...,\gamma _d } \right)\,} {\rm{d}}\gamma _1 ...{\rm{d}}\gamma _d 
\]
and
\[
p\left( {\bf{x}} \right) = \int\limits_{\rho _1  - {\rm{i}}\infty }^{\rho _1  + {\rm{i}}\infty } {...} \int\limits_{\rho _d  - {\rm{i}}\infty }^{\rho _d  + {\rm{i}}\infty } {G_p \left( {{\bf{x}},\gamma _1 ,...,\gamma _d } \right){\rm{d}}\gamma _1 ...{\rm{d}}\gamma _d } 
\]

The former can now be approximated by the positions: $\gamma _j  = \rho _j  + {{\rm{i}}} \eta _j $, $ \eta _j= r_j \Delta _j $, with $-m \le r_j\le m$, $m \in \mathbb{N}$ and $\Delta _j \in \mathbb{R}$. Extremes of integration have been truncated up to $\left[ { - m\Delta _j ,m\Delta _j } \right]$. The approximated dual series representing the joint statistics reads

\[
\varphi \left( {\bf{u}} \right) = {\rm{i}}^d \sum\limits_{r_1  =  - \infty }^\infty  {...\sum\limits_{r_d  =  - \infty }^\infty  {G_\varphi  \left( {{\bf{u}},\rho _1  + {\rm{i}}r_1 \Delta _1 ,...,\rho _d  + {\rm{i}}r_d \Delta _d } \right)\,} } \Delta _1 ...\Delta _d 
\]

\[
p\left( {\bf{x}} \right) = {\rm{i}}^d \sum\limits_{r_1  =  - \infty }^\infty  {...\sum\limits_{r_d  =  - \infty }^\infty  {G_p \left( {{\bf{x}},\rho _1  + {\rm{i}}r_1 \Delta _1 ,...,\rho _d  + {\rm{i}}r_d \Delta _d } \right)\,} } \Delta _1 ...\Delta _d 
\]

Moreover, the presence of the gamma function in the integrand of eqs.(\ref{equ29}) and (\ref{equ32}) ensures that the integrand vanishes and, consequently, the approximated sums can be truncated in the form 

\[
\varphi \left( {\bf{u}} \right) = {\rm{i}}^d \sum\limits_{r_1  =  - m_1 }^{m_1 } {...\sum\limits_{r_d  =  - m_d }^{m_d } {G_\varphi  \left( {{\bf{u}},\rho _1  + {\rm{i}}r_1 \Delta _1 ,...,\rho _d  + {\rm{i}}r_d \Delta _d } \right)\,} } \Delta _1 ...\Delta _d 
\]

\[
p\left( {\bf{x}} \right) = {\rm{i}}^d \sum\limits_{r_1  =  - m_1 }^{m_1 } {...\sum\limits_{r_d  =  - m_d }^{m_d } {G_p \left( {{\bf{x}},\rho _1  + {\rm{i}}r_1 \Delta _1 ,...,\rho _d  + {\rm{i}}r_d \Delta _d } \right)\,\Delta _1 ...\Delta _d \,} } 
\]
with $m_j  \in \mathbb{N}$. In this way, the accuracy of the results is of course affected by discretization and a truncation error. By means of some applications we show that the truncated sums are nevertheless very good approximations, computing a finite number of complex moments.

\subsubsection{Bivariate Cauchy random vector}
\begin{figure}[b]
\includegraphics[width=\columnwidth]{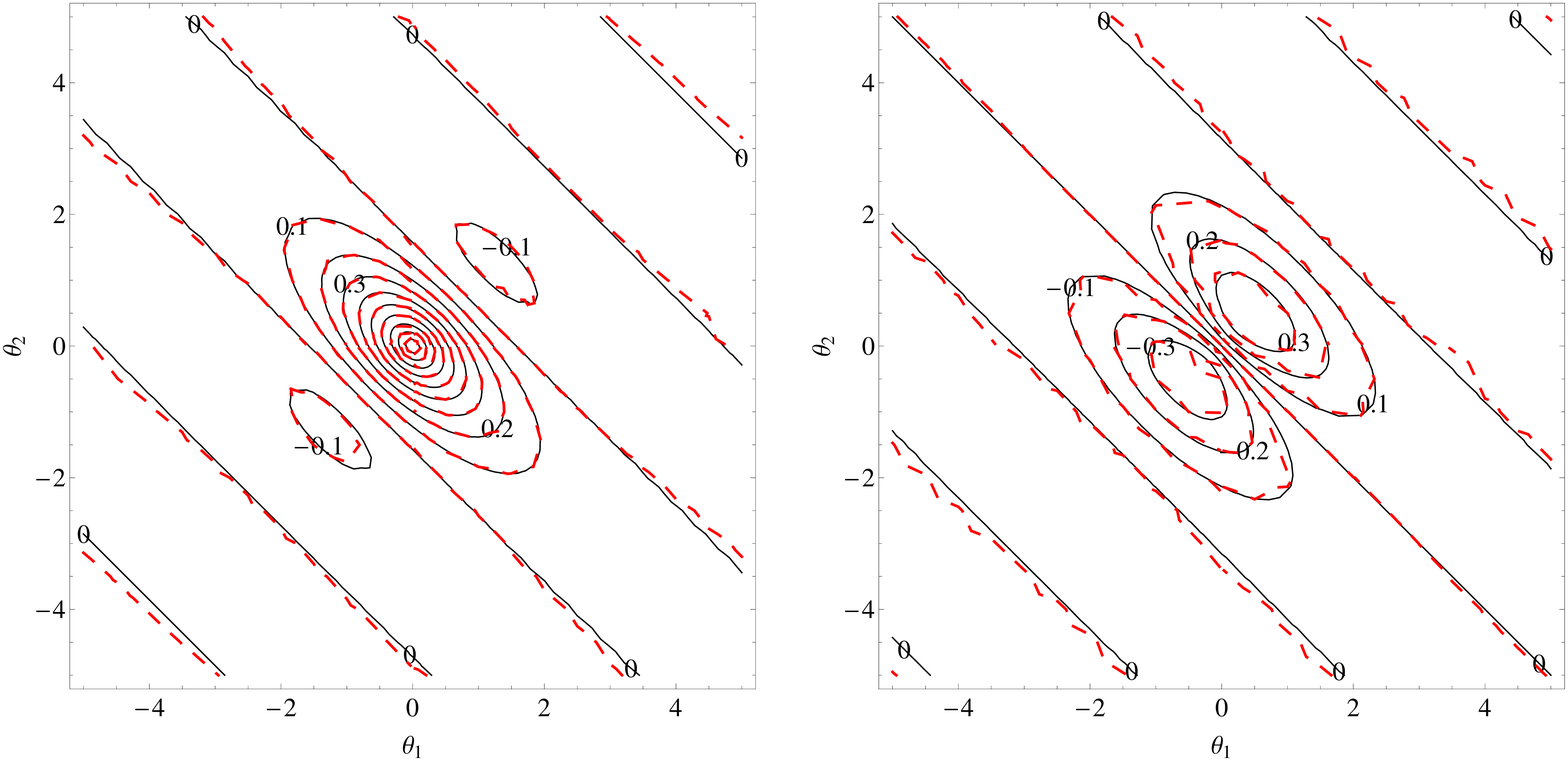}
\caption{Comparison between the exact (continuous) and the approximated (dashed) real part of the joint characteristic function (left panel) and the imaginary part of the joint characteristic function(right panel) of bivariate Cauchy distribution}
\label{fig4ab}
\end{figure}

\begin{figure}[b]
\centering
\includegraphics[scale=1]{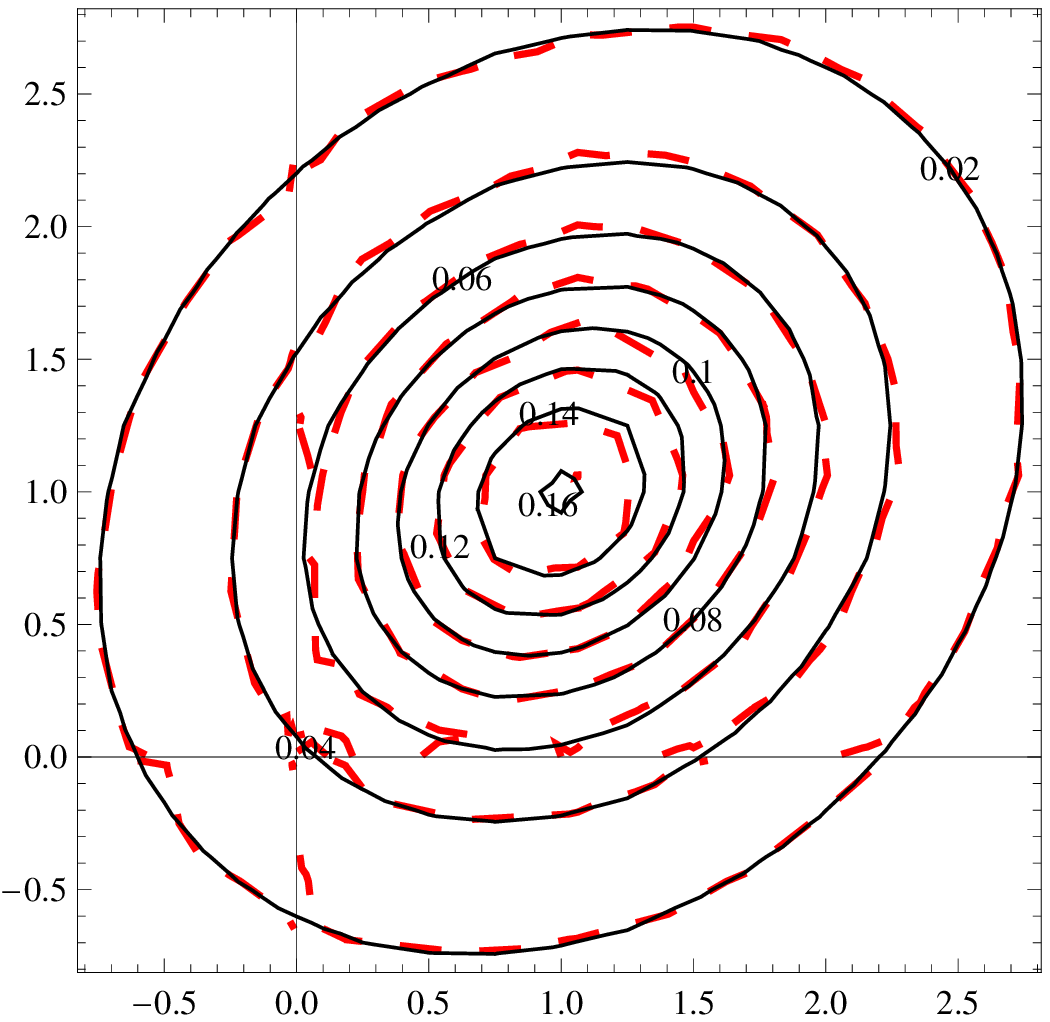}
\caption{Comparison between the exact (continuous) and the approximated (dashed) joint PDF of bivariate Cauchy distribution with no-zero location}
\label{fig5}
\end{figure}

In this example we deal with the bivariate Cauchy distribution, that is an $\alpha$-stable distribution with $\alpha=1$. Then, it has power-law tails and, as pointed out already in the introduction it has not integer moments of any order. The expression of its joint PDF is in general
\[
p\left( \bm{x} \right) = \frac{{c\det \left( {\bm{\Sigma }} \right)^{ - 1/2} }}{{\left( {1 + \left( {{\bf{x}} - {\bm{\mu }}} \right)^T {\bm{\Sigma }}^{ - 1} \left( {{\bf{x}} - {\bm{\mu }}} \right)} \right)^{3/2} }}
\]
while its joint characteristic function has the form
\[
\varphi \left( {\bf{u}} \right) = \exp \left( { - {\bf{u}}^T {\bm{\Sigma u}}} \right)^{1/2}  + {\rm i}{\bf{u}}^T {\bm{\mu }}
\]
and $c = \pi ^{ - 3/2} \Gamma \left( {3/2} \right)$. The matrix ${\bm{\Sigma}}$ and the ${\bm{\mu}}$ 
have the meaning of scale matrix and location vector, in some sense analogous to the covariance matrix and the mean vector. For the application of the fractional moments approach, we consider in this example the parameters 
\[{\bm{ \Sigma}}  = \left( {\begin{array}{*{20}c}
   1 & {0.2}  \\
   {0.2} & 1  \\
\end{array}} \right);\,\,{\bm{\mu} } = \left( {\begin{array}{*{20}c}
   1  \\
   1  \\
\end{array}} \right)
\]
then, as the location vector is not zero, we will expect a density with the maximum in the point $x_1= 1$ and $x_2=1$. Consequently, due to the condition ${\bm{\mu}}\ne {\bf{0}}$, the joint characteristic function of such random vector is a complex function as plainly descends from the definition. Then, eq.(\ref{equ29}) and eq.(\ref{equ32}) are discretized by  choosing the steps $\Delta_1  = \Delta_2  = 0.3$, and the integration is performed in the fundamental strip with $\rho _1  = \rho _2  = 0.5$. By using $m=20$ fractional moments we show that very good numerical convergence is obtained. Indeed, in Fig.\ref{fig4ab} we first give the contour plots of the real and the imaginary part of the complex characteristic function where the dashed line is the approximation proposed. Without performing any further Fourier transform but just applying the discrete form of eq.(\ref{equ32}), with the same $20$ fractional moments the joint PDF is calculated and plotted in Fig.\ref{fig5} showing a good numerical convergence in the whole domain.

\section{Conclusions}
\label{sec:Conclusions}

\noindent In this paper, we provide new insights into the problem of the statistical representation of random variables by means of the Mellin transform and fractional calculus. We define a class of complex moments (comprehending fractional moments) that can represent every kind of joint PDF and joint characteristic function, including joint characteristic functions that are non-differentiable in zero, and densities with heavy-tails, like the bivariate Cauchy distribution or other power-law distributions. Comparison with classical applications of inverse Mellin transform shows that starting from the Mellin transform of the characteristic function and exploiting its Hermitian nature an easy and efficient representation of the statistics of random variables and vectors is achieved. Further, the complex moments introduced represent both the joint PDF and the joint CF. 

The perspective presented in this paper might open, to the authors' opinion, a new touch in the moment problem, that is, under which conditions the knowledge of moments suffices to reconstruct the underlying density, to hierarchy of moments and related issues.

\bibliographystyle{elsarticle-num}
\bibliography{biblio} 

\begin{thebibliography}{10}
\expandafter\ifx\csname url\endcsname\relax
  \def\url#1{\texttt{#1}}\fi
\expandafter\ifx\csname urlprefix\endcsname\relax\def\urlprefix{URL }\fi
\expandafter\ifx\csname href\endcsname\relax
  \def\href#1#2{#2} \def\path#1{#1}\fi

\bibitem{metz00}
R.~Metzler, J.~Klafter, The random walk's guide to anomalous diffusion: A
  fractional dynamics approach, Phys. Rep 339 (2000) 1--77.

\bibitem{metz04}
R.~Metzler, J.~Klafter, The restaurant at the end of the random walk: Recent
  developments in the description of anomalous transport by fractional
  dynamics, Journal of Physics A: Mathematical and General 37 (2004)
  R161--R208.

\bibitem{brockmann}
D.~Brockmann, L.~Hufnagel, T.~Geisel, The scaling laws of human travel, Nature
  439~(7075) (2006) 462--465.

\bibitem{barabasi}
M.~Gonzalez, C.~Hidalgo, A.~Barabasi, Understanding individual human mobility
  patterns, Nature 453~(7196) (2008) 779--782.

\bibitem{lomholt}
T.~Lomholt, M.A.and~Koren, R.~Metzler, J.~Klafter, L\'{e}vy strategies in
  intermittent search processes are advantageous, Proceedings of the National
  Academy of Sciences of the United States of America 105~(32) (2008)
  11055--11059.

\bibitem{mateos}
G.~Ramos-Fernandez, J.~Mateos, O.~Miramontes, G.~Cocho, H.~Larralde,
  B.~Ayala-Orozco, L\'{e}vy walk patterns in the foraging movements of spider
  monkeys (ateles geoffroyi), Behavioral Ecology and Sociobiology 55~(3) (2004)
  223--230.

\bibitem{chechkin}
A.~Chechkin, V.~Gonchar, M.~Szydlowski, Fractional kinetics for relaxation and
  superdiffusion in a magnetic field, Physics of Plasmas 9~(1) (2002) 78.

\bibitem{diego}
D.~Del-Castillo-Negrete, Chaotic transport in zonal flows in analogous
  geophysical and plasma systems, Physics of Plasmas 7~(5 II) (2000)
  1702--1711.

\bibitem{duplantierreview}
B.~Duplantier, Statistical mechanics of polymer networks of any topology,
  Journal of Statistical Physics 54~(3-4) (1989) 581--680.

\bibitem{lomholt1}
M.~Lomholt, T.~Ambj\"{o}rnsson, R.~Metzler, Optimal target search on a
  fast-folding polymer chain with volume exchange, Physical Review Letters
  95~(26) (2005) 1--4.

\bibitem{scher}
H.~Scher, E.~Montroll, Anomalous transit-time dispersion in amorphous solids,
  Physical Review B 12~(6) (1975) 2455--2477.

\bibitem{scher1}
H.~Scher, G.~Margolin, R.~Metzler, J.~Klafter, B.~Berkowitz, The dynamical
  foundation of fractal stream chemistry: The origin of extremely long
  retention times, Geophys. Res. Lett. 29, 1061~(5) (2002) 1061.
\newblock \href {http://dx.doi.org/10.1029/2001GL014123}
  {\path{doi:10.1029/2001GL014123}}.

\bibitem{weeks}
T.~Solomon, E.~Weeks, H.~Swinney, Observation of anomalous diffusion and
  l\'{e}vy flights in a two-dimensional rotating flow, Physical Review Letters
  71~(24).

\bibitem{gzyl06}
H.~Gzyl, P.~N. Inverardi, A.~Tagliani, M.~Villasana, Maxentropic solution of
  fractional moment problems, Appl. Math. Comput. 173 (2006) 109--205.

\bibitem{novi03}
P.~N. Inverardi, A.~Petri, G.~Pontuale, A.~Tagliani, Hausdorff moment problem
  via fractional moments, Applied Stochastic Models in Business and Industry
  144 (2003) 61--74.

\bibitem{novi05}
P.~N. Inverardi, A.~Petri, G.~Pontuale, A.~Tagliani, Stieltjes moment problem
  via fractional moments, Appl. Math. Comput. 166 (2005) 664--677.

\bibitem{nigm09}
R.~Nigmatullin, D.~Baleanu, A.~{Din\c{c} E Solak}, Characterization of a
  benzoic acid modified glassy carbon electrode expressed quantitatively by new
  statistical parameters, Physica E: Low-dimensional and Nanostructures 41
  (2005) 609--616.

\bibitem{epst48}
B.~Epstein, Some applications of the mellin transform in statistics, The Annals
  of Mathematical Statistics 19~(3) (1948) 370--379.

\bibitem{hunt39}
E.~V. Huntington, Frequency distribution of product and quotient, The Annals of
  Mathematical Statistics 10~(2) (1939) 195--198.

\bibitem{dola64}
B.~Dolan, The mellin transform for moment–generation and for the probability
  density of products and quotients of random variables, Proceedings of the
  IEEE 52~(12) (1964) 1745--1746.

\bibitem{spri79}
M.~Springer, The Algebra of Random Variables, Wiley Series in Probability and
  Mathematical Statistics, 1979.

\bibitem{math93b}
A.~Mathai, A Handbook of Generalized Special Functions for Statistical and
  Physical Sciences, Oxford University Press, 1979.

\bibitem{math08}
A.~Mathai, H.~Haubold, Special Functions for Applied Scientists, Springer,
  2008.

\bibitem{math78}
A.~Mathai, R.~Saxena, The H-function with Applications in Statistics and
  Physical Science, Wiley Halsted, New York, 1978.

\bibitem{math93}
A.~M. Mathai, The residual effect of a growth-decay mechanism and the
  distributions of covariance structures, The Canadian Journal of Statistics /
  La Revue Canadienne de Statistique 21~(3) (1993) 277--283.

\bibitem{prov86}
S.~Provost, The exact distribution of the ratio of a linear combination of
  chi-square variables over the root of a product of chi-square variables, The
  Canadian Journal of Statistics 14 (1986) 61--67.

\bibitem{prov88}
S.~Provost, The exact density of a general linear combination of gamma
  variates, Metron 46 (1988) 61--69.

\bibitem{wolf75}
S.~Wolfe, On moments of probability distribution function, in: B.~Ross (Ed.),
  Fractional Calculus and Its Applications, Springer-Verlag, Berlin, 1975.

\bibitem{wolf71}
S.~J. Wolfe, On moments of infinitely divisible distribution functions, The
  Annals of Mathematical Statistics 42~(6) (1971) 2036--2043.

\bibitem{wolf73}
S.~J. Wolfe, On the local behavior of characteristic functions, The Annals of
  Probability 1~(5) (1973) 862--866.

\bibitem{wolf75b}
S.~J. Wolfe, On derivatives of characteristic functions, The Annals of
  Probability 3~(4) (1975) 737--738.

\bibitem{laue80}
G.~Laue, Remarks on the relation between fractional moments and fractional
  derivatives of characteristic functions, Journal of Applied Probability
  17~(2) (1980) 456--466.

\bibitem{laue83}
G.~Laue, Remarks on the theory of characteristic functions of non-negative
  random variables, Sankhya-: The Indian Journal of Statistics, Series A 45~(1)
  (1983) 44--55.

\bibitem{laue86}
G.~Laue, Results on moments of non-negative random variables, Sankhya-: The
  Indian Journal of Statistics, Series A 48~(3) (1986) 299--314.

\bibitem{cott09}
G.~Cottone, M.~{Di Paola}, On the use of fractional calculus for the
  probabilistic characterization of random variables, Probabilistic Engineering
  Mechanics 24 (2009) 321--330.

\bibitem{cott09b}
G.~Cottone, M.~{Di Paola}, A new representation of power spectral density and
  correlation function by means of fractional spectral moments, Probabilistic
  Engineering Mechanics.

\bibitem{cott09c}
G.~Cottone, M.~{Di Paola}, R.~Santoro, A novel exact representation of
  stationary colored gaussian processes (fractional differential approach),
  Journal of Physics A: Mathematical and Theoretical.

\bibitem{samk93}
S.~Samko, A.~Kilbas, O.~Marichev, Fractional Integrals and Derivatives. Theory
  and Applications, Gordon and Breach Science Publishers, Switzerland, 1993.

\bibitem{sriv95}
H.~Srivastava, R.~Saxena, J.~Ram, Some multidimensional fractional integral
  operators involving a general class of polynomials, Journal of mathematical
  analysis and applications 193 (1995) 373--389.

\bibitem{samo94}
G.~Samorodnitsky, M.~Taqqu, Stable non-gaussian random processes, Chapman and
  Hall/CRC, 1994.

\end{thebibliography}

\end{document}